%% file: main.tex
\documentclass[%
 amsmath,amssymb,
 aps,
]{revtex4-1}

\usepackage{graphicx}
\usepackage{dcolumn}
\usepackage{bm}

\begin{document}


\title{Estimating the Anisotropy of Protein Structures from SAXS}

\author{Biel Roig-Solvas}
\email[Corresponding Author: ]{biel@ece.neu.edu}
\author{Dana H. Brooks}%

\affiliation{%
 Department of Electrical and Computer Engineering, Northeastern University, Boston MA\\}
 \author{Lee Makowski}%
 \affiliation{
 Department of Bioengineering, Northeastern University, Boston MA\\}%

\begin{abstract}
\input{Tex/Abstract.tex}
\end{abstract}

\maketitle


\section{Introduction}\label{Section:Intro}
\input{Tex/Introduction.tex}
\section{Methods}\label{Section:Methods}
\input{Tex/Methods.tex}
\section{Results and Discussion}\label{Section:Results}
\input{Tex/Results.tex}
\section{Conclusions}\label{Section:Conclusions}
\input{Tex/Conclusions.tex}
\section*{Acknowledgements}
The authors would like to acknowledge Hao Zhou, Jenifer Winters and John Badger for data collection and fruitful discussions. This work was supported by the U.S. Department of Energy, Office of Science, Basic Energy Sciences (grant no. DE–SC0000997).
 
 \bibliographystyle{unsrt}
 \bibliography{bibliography}

\end{document}

%% file: Tex/Abstract.tex
In the field of small angle x-ray scattering (SAXS), the task of estimating the size of particles in solution is usually synonymous with the Guinier plot. The approximation behind this plot, developed by Guinier in 1939 
provides a simple yet accurate characterization of the scattering behavior of particles at low scattering angle $q$, together with a computationally efficient way of inferring their radii of gyration $R_G$. Moreover, this approximation is valid beyond spherical scatterers, making its use ubiquitous in the SAXS world. However, when it is important to estimate further particle characteristics, such as the anisotropy of the scatterer's shape, no similar or extended approximations are available. Existing tools to characterize the shape of scatterers rely either on prior knowledge of the scatterers' geometry or on iterative procedures to infer the particle shape \textit{ab initio}.\\
\\
 In this work we develop a low angle approximation of the scattering intensity $I(q)$ for ellipsoids of revolution and show how to extract size and anisotropy information from the parameters of that approximation. Beyond ideal ellipsoids of revolution, we show that this approximation can be used to infer the size and shape of molecules in solution, both in computational and experimental scenarios. We discuss the limits of our approach and study the impact of a particle's anisotropy in the Guinier estimate of $R_G$.

%% file: Tex/Introduction.tex
The Guinier approximation \cite{Guinier1939} is a ubiquitous tool to assess the size of scatterers in SAXS experiments. This approximation relates the radius of gyration $R_G$ of a particle to the exponential decay of the scattering curve $I(q)$ around the origin:
\begin{equation}
    I(q) \approx I(0)\;e^{\frac{-q^2\;R_G^2}{3}}\quad \text{for}\quad q\,R_G \leq 1.3
    \label{GuinierAppox}
\end{equation}
where $R_G$ is measured as the linear decay of the plot $ln(I(q))$ vs. $q^2$, commonly known as the Guinier plot \cite{Putnam2007}. As the parameter $R_G$ is model-independent, this approximation can be applied to any type of scatterer, which explains the Guinier approximation's universal presence in the SAXS community.\\ 
\\
Additional features of a scatterer, like its anisotropy, can be recovered from $I(q)$ if one knows a priori the geometry of the scatterer. For the case of simple shapes (e.g. cylinders, parallelepipeds, ellipsoids, etc.), the curve $I(q)$ has a clear relationship with the geometric parameters \cite{Feigin1989}, and these parameters can be estimated by iterative curve fitting \cite{Konarev2003}. However, most biomolecules deviate from this ideal geometry, and ab initio modeling is typically applied in order to estimate their shape \cite{Franke2009}.\\
\\
In this work we show that one can approximate the anisotropy of biological scatterers by assuming an ellipsoid-like behavior of $I(q)$ for low $q$. To do so, we develop a fourth-order approximation of the scattering curve $I(q)$ for ellipsoids of revolution and present a Guinier-like parameter estimation to infer both size and anisotropy from the curve $I(q)$ in a direct, non-iterative manner. These estimates can be used to gather fast estimates of the approximate particle size and anisotropy and to get better initializations for ab initio iterative methods.\\
\\
We test the presented approach with both computed $I(q)$ curves from molecules in the PDB database \cite{Berman2000} and real data from the BIOSIS database \cite{Hura2009} and compare the estimated geometries with their molecular structure. The presented method successfully approximates the anisotropy for all the test cases, showing its applicability for SAXS data analysis.\\
\\
The paper is structured as follows: we start Section 2 by reviewing the results in the literature regarding the x-ray scattering of ellipsoidal bodies. From those results, we derive an approximation of $I(q)$ as a function of the ellipsoid's semiaxes. A simple least squares algorithm is proposed to carry out a polynomial fit to the $I(q)$ curve and to recover the semiaxes from the parameters of that fit. In Section 3 we numerically validate the proposed approach, by computing the $I(q)$ curve of a series of ideal ellipsoids of revolution across a large aspect ratio range and then recover the parameters of those ellipsoids using the proposed approach. This experiment is then extended to molecular models from the PDB database, whose scattering is computed using the FoXS software \cite{Schneidman-Duhovny2010}, and to 6 experimental SAXS datasets from the BIOISIS database for which a computational model of the scatterer was available. At the end of Section 3 we discuss the accuracy of $R_G$ estimation between the proposed approximation and the Guinier plot and we summarize our conclusions in Section 4. 

%% file: Tex/Methods.tex
\subsection{Background}
The scattering behavior of ellipsoids of revolution has been widely studied in the SAXS field, going back to Guinier's seminal work in X-ray scattering from 1939 \cite{Guinier1939}. As also noted in the same year by Patterson \cite{Patterson1939}, the form factor for this class of ellipsoids, with semiaxes $R,R,\epsilon R$, where $\epsilon$ is the axial ratio of the ellipsoid, can be expressed as:
\begin{equation}
    S\left(u\right) = 9\left(\frac{\sin\left(u\right) - u\,\cos\left(u\right)}{\left(u\right)^3}\right)^2
    \label{Eq:S_of_u}\quad \text{for}\quad u = q\,R\left(\theta,\phi\right) \quad \text{and} \quad R\left(\theta,\phi\right)  = R\sqrt{\sin^2\left(\phi\right) + \epsilon^2\,\cos^2\left(\phi\right)}
\end{equation}
where $q \in [0,\infty)$, $\theta \in [0,2\pi]$ and $\phi \in [0,\pi]$
are the radius and azymuthal and inclination angle of the spherical coordinates in reciprocal space. The SAXS intensity of the ellipsoid is obtained by averaging equation \ref{Eq:S_of_u} across all possible all orientations, which yields:
\begin{equation}
    I_{R}(q) = \frac{1}{4\pi} \; \int_{\theta=0}^{2\pi}\; \int_{\phi=0}^{\pi}\; S\left(q\,R\left(\theta,\phi\right)\right)\,\sin\left(\phi\right)\,d\theta\,d\phi
    \label{Eq:I(q)_Integral}
\end{equation}
This integral has the structure of a hypergeometric function \cite{Roess1947}, which can only be evaluated numerically. Due to this lack of a closed analytic form, the ellipsoid parameters $R$ and $\epsilon$ cannot be estimated directly from the $I_{R}(q)$ curve, so methods that rely either on iterative curve fitting \cite{Sholer1975,Konarev2003} or comparison with precomputed curves \cite{Feigin1989} have been proposed to solve this estimation problem. In this work we propose to carry out an approximation of $I_{R}(q)$ around the origin and use the coefficients of that approximation to estimate the ellipsoid parameters.

\subsection{Approximating $I_{R}(q)$}
The first step in the approximation is to express the function $S\left(u\right)$ in equation \ref{Eq:S_of_u} as a Taylor series. Expanding the square we get:
\begin{equation}
    S(u) = 9\left( \frac{\sin^2(u)}{u^6} + \frac{\cos^2(u)}{u^4} - \frac{2\sin(u)\cos(u)}{u^5} \right)
\end{equation}
Expressing these three terms as Taylor series and taking the negative powers of $u$ outside of the summation one gets:
\begin{equation}
    \frac{\sin^2(u)}{u^6} = \frac{1}{u^4}-\frac{1}{3u^2}+\frac{1}{2\,u^6}\sum_{m=3}^\infty (-1)^m \frac{(2u)^{2m}}{(2m)!}
    \label{Eq:Taylor1}
\end{equation}
\begin{equation}
    \frac{\cos^2(u)}{u^4} = \frac{1}{u^4}-\frac{1}{u^2}+\frac{1}{2\,u^4}\sum_{m=2}^\infty (-1)^m \frac{(2u)^{2m}}{(2m)!}
    \label{Eq:Taylor2}
\end{equation}
\begin{equation}
    \frac{-2\sin(u)\cos(u)}{x^5} =  \frac{-2}{u^4}+\frac{4}{3u^2}+\frac{1}{u^5}\sum_{m=3}^\infty (-1)^m \frac{(2u)^{2m-1}}{(2m-1)!}
    \label{Eq:Taylor3}
\end{equation}
Applying the change of indices $n = m-3$, $n = m-2$ and $n = m-3$ in equations \ref{Eq:Taylor1}, \ref{Eq:Taylor2} and \ref{Eq:Taylor3}, respectively, and adding the three terms together, we get a single summation in the form of:
\begin{equation}
\label{Eq:S_Taylor}
\begin{split}
    S(u) &= 9 \sum_{n=0}^\infty (-1)^n\,(2u)^n\,\left(\frac{2^5}{(2n+5)!}+\frac{2^3}{(2n+4)!}+\frac{2^5}{(2n+6)!} \right) \\
    &=  9\,\sum_{n=0}^\infty\,\left(-1\right)^n\,2^4\,\frac{\left(2\,n+5\right)\left(n+1\right)}{\left(2\,n +6\right)!} \left(2\,u\right)^{2\,n}
\end{split}
\end{equation}
Adding this final expression into the $I_{R}(q)$ integral in equation \ref{Eq:I(q)_Integral} and exchanging the order of the integral and the summation, we get:
\begin{equation}
\begin{split}
     I_{R}(q) &= \frac{9}{4\pi} \; \sum_{n=0}^\infty\,\left(-1\right)^n\,2^4\,\frac{\left(2\,n+5\right)\left(n+1\right)}{\left(2\,n +6\right)!}\, \int_{\theta=0}^{2\pi}\;\int_{\phi=0}^{\pi}\; \left(2\,q\,R\left(\theta,\phi\right)\right)^{2\,n}\,\sin\left(\phi\right)\,d\theta\,d\phi\\
     &= \frac{9}{4\pi} \; \sum_{n=0}^\infty\,\left(-1\right)^n\,2^4\,\frac{\left(2\,n+5\right)\left(n+1\right)}{\left(2\,n +6\right)!}\, (2\,qR)^{2n}\,  \int_{\theta=0}^{2\pi}\;\int_{\phi=0}^{\pi}\; \left(\sin^2\left(\phi\right) + \epsilon^2\,\cos^2\left(\phi\right)\right)^{n}\,\sin\left(\phi\right)\,d\theta\,d\phi\\
     \end{split}
    \label{Eq:I(q)_Taylor}
\end{equation}
Next we rewrite the double integral above into a finite series as a function of powers of $\epsilon$. First we note that no term inside the integral depends on $\theta$, so that integral reduces to a $2\pi$ constant factor. Using the equality $\sin^2\left(\phi\right) = 1 - \cos^2\left(\phi\right)$ and applying the change of variables $x = \cos\left(\phi\right)$, we get:
\begin{equation}
\begin{split}
 \int_{\theta=0}^{2\pi}\;\int_{\phi=0}^{\pi}\; \left(\sin^2\left(\phi\right) + \epsilon^2\,\cos^2\left(\phi\right)\right)^{n}\,\sin\left(\phi\right)\,d\theta\,d\phi\,&=\,-2\pi\, \int_{x=1}^{-1}\; \left(1 - \left(1-\epsilon^2\right)\,x^2\right)^n\,dx\\
    &=\,4\pi\,\int_{x=0}^{1}\; \left(1 - \left(1-\epsilon^2\right)\,x^2\right)^n\,dx
\end{split}
\end{equation}
Finally, we apply the binomial theorem to the term inside the integral and get:
\begin{equation}
    \begin{split}
       4\pi\,\int_{x=0}^{1}\; \left(1 - \left(1-\epsilon^2\right)\,x^2\right)^n\,dx\,&=\,4\pi\,\sum_{k=0}^n {n \choose k} \left(\epsilon^2-1\right)^k\,\int_{x=0}^{1}\,x^{2k}\,dx \\
        &=\,4\pi\,\sum_{k=0}^n {n \choose k} \frac{\left(\epsilon^2-1\right)^k}{2k+1}
    \end{split}
\end{equation}
which leads to the infinite series:
\begin{equation}
     I_{R}(q) = 9\sum_{n=0}^\infty\,\left(-1\right)^n2^4\frac{\left(2n+5\right)\left(n+1\right)}{\left(2n +6\right)!}(2\,qR)^{2n}\sum_{k=0}^n {n \choose k} \frac{\left(\epsilon^2-1\right)^k}{2k+1}
     \label{Eq:I(q)_series}
\end{equation}
Developing the first three terms of this series, we conclude:
\begin{equation}
    \begin{split}
    I_{R}(q) = 9&( \;\frac{2^4\,5}{6!}
     - \frac{2^6\,14}{8!}\,(qR)^{2}\,\left(1 + \frac{\left(\epsilon^2-1\right)}{3} \right)
      + \frac{2^8\,27}{10!}\,(qR)^{4}\,\left(1 + \frac{2\left(\epsilon^2-1\right)}{3} +\frac{\left(\epsilon^2-1\right)^2}{5}\right) ) +\,\mathcal{O}(q^6) \\
    =\;&\; 1 - \frac{R^2 (2+ \epsilon^2)}{15}\, q^2 + \frac{1}{21} \left( \left(\frac{R^2 (2+ \epsilon^2)}{5} \right)^2 + \frac{1}{5} \left(\frac{R^2(2-2\epsilon^2) }{5}\right)^2 \right)\, q^4 +\,\mathcal{O}(q^6)
    \end{split}
    \label{AlmostFinalApprox}
\end{equation}
Using the expression for the radius of gyration of an ellipsoid of revolution $R_G^2 = \frac{R^2 +R^2 + \epsilon^2R^2}{5}$, we note that the quadratic term in the Equation \ref{AlmostFinalApprox} corresponds to $\frac{-R_G^2}{3}$, as in Equation \ref{GuinierAppox}. The quartic term depends on the fourth power of the radius of gyration and an additional term which we refer to as the anisotropy factor $ A_F^2 = \left(\frac{R^2(2-2\epsilon^2) }{5}\right)^2$. This anisotropy factor is 0 for the spherical case $\epsilon = 1$, and increases as $\epsilon$ deviates from 1, i.e. as the anisotropy of the ellipsoid increases. Adding these terms to Equation \ref{AlmostFinalApprox} we get the final form of the approximation:
\begin{equation}
     I_{R}(q) = \; 1 - \frac{R_G^2}{3}\, q^2 + \frac{1}{21} \left( R_G^4 + \frac{1}{5}  A_F^2 \right)\, q^4 +\,\mathcal{O}(q^6)
     \label{FinalApprox}
\end{equation}
\\
One can estimate the parameters $R_G^2$ and $A_F^2$ by performing a simple least squares polynomial fit of order $P$ on the scattering data $I(q)$, i.e. find the parameters $[a_0, a_1, \dots, a_P]$ such that the norm of the error $\xi = I(q) - \sum_i^P a_i q^{2i}$ is minimized. Calculating $R$ and $\epsilon$ from the estimated $R_G^2$ and $A_F^2$ leads to two solutions, one for the oblate case $\epsilon<1$ and one for the prolate case $\epsilon \geq 1$. In low noise situations, one can decide between these cases by computing the $q^6$ term in \ref{Eq:I(q)_series} for both pairs of $R$ and $\epsilon$ and comparing those to the estimated $q^6$ term in the polynomial fit. However, higher order terms are increasingly sensitive to small variations of the ellipsoid parameters, so this approach might not work in most experimental scenarios, where the presence of noise can perturb the estimation of the ellipsoid parameters $R$ and $\epsilon$. In that case, the proposed approach won't be able to disambiguate between the oblate and prolate solutions.  \\
\\

%% file: Tex/Results.tex
In this section we put to practice the method presented in the previous section. First, we test the approach with scattering curves from ideal ellipsoids, scanning through a wide range of aspect ratios. Then we apply the proposed method to SAXS curves from different molecules, reporting on both curves calculated from computational models and data measured in 
SAXS experiments.\\


\subsection{Anisotropy estimation for ideal ellipsoids}
\begin{figure}
\includegraphics[width=18cm]{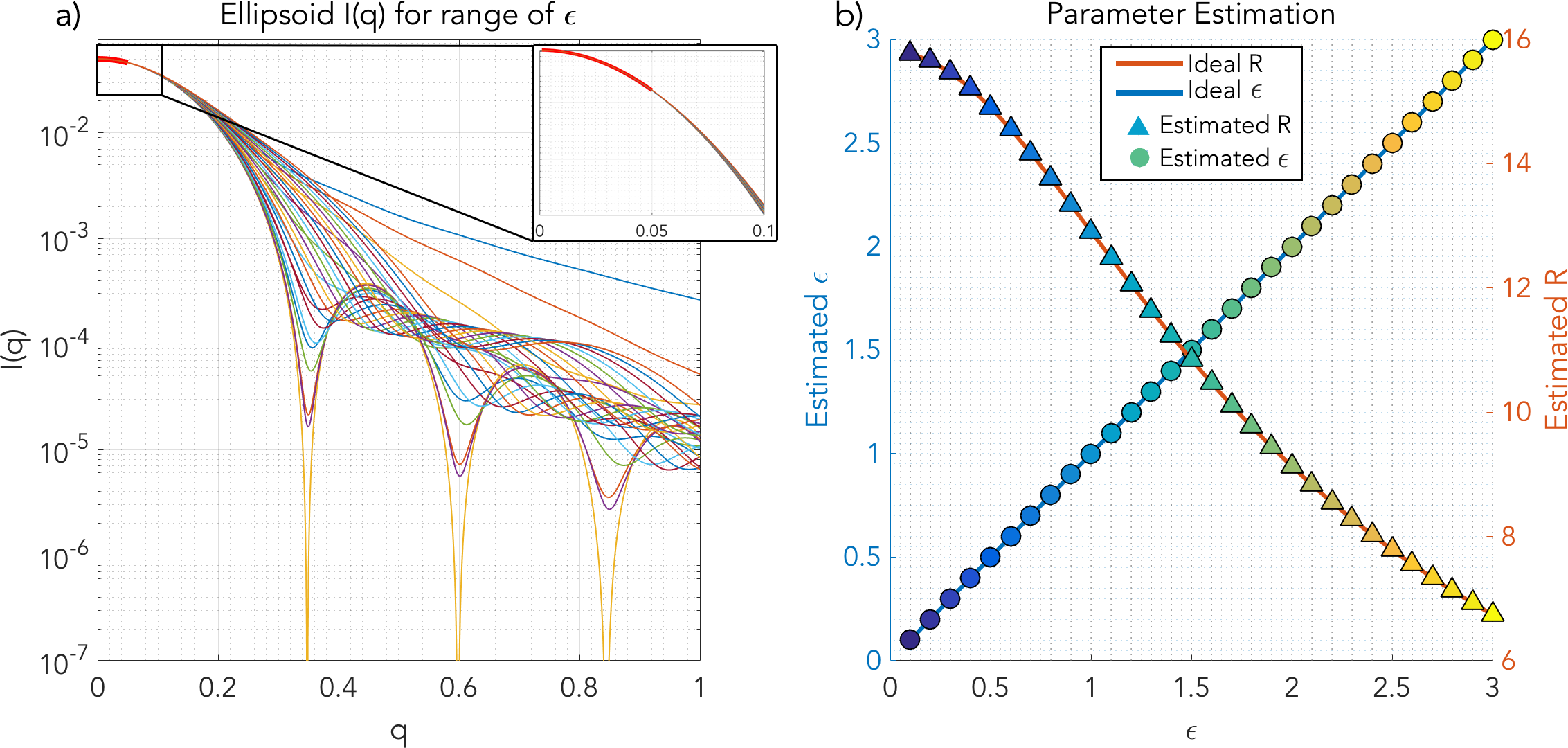}
\caption{a) $I_R(q)$ curves for all ellipsoids in the chosen $\epsilon$ anisotropy range. The red segment, amplified in the inset, shows the interval over which the fitting was carried out. b) Parameter estimation results per each test case. The continuous blue and red curves indicate the expected $\epsilon$ and $R$ values as a function of $\epsilon$, while circles and triangles indicate the estimated $\epsilon$ and $R$, respectively, for each test case.}
\label{Fig:EllipsoidsIq}
\end{figure}

To test the proposed parameter estimation in the ideal case, we computed a set of scattering curves $I_R(q)$ by numerically evaluating the integral in Equation \ref{Eq:I(q)_Integral}. To cover a wide set of test cases, we fixed $R_G$ of the ellipsoids to 10 and scanned the range of 30 $\epsilon$ values from 0.1 to 3. The corresponding curves can be seen in Figure \ref{Fig:EllipsoidsIq}.a.\\

To estimate the parameters, a least squares polynomial fit was performed on the computed $I_R(q)$ curves in the range $0<q<0.05$, as shown in the inset of Figure \ref{Fig:EllipsoidsIq}.a. A polynomial order of $P = 6$ was used for the fit and the oblate/prolate candidate was chosen by selecting the one that better approximated the $q^6$ fitting term, as explained in Section 2. The results of the parameter estimation are shown in Figure \ref{Fig:EllipsoidsIq}.b for the test cases studied. The solid lines show the actual values of $\epsilon$ (left axis) and $R$ (right axis) as a function of $\epsilon$ for an ellipsoid of revolution with $R_G = 10$. For each test case, the estimated $\epsilon$ and $R$ are plotted in circles and triangles, respectively. As shown by the agreement of the estimated parameters with the ideal curves, the proposed method is able to correctly estimate both $\epsilon$ and $R$ for all test cases.\\


\subsection{Anisotropy estimation for molecules}

\begin{figure}
\includegraphics[width=15cm]{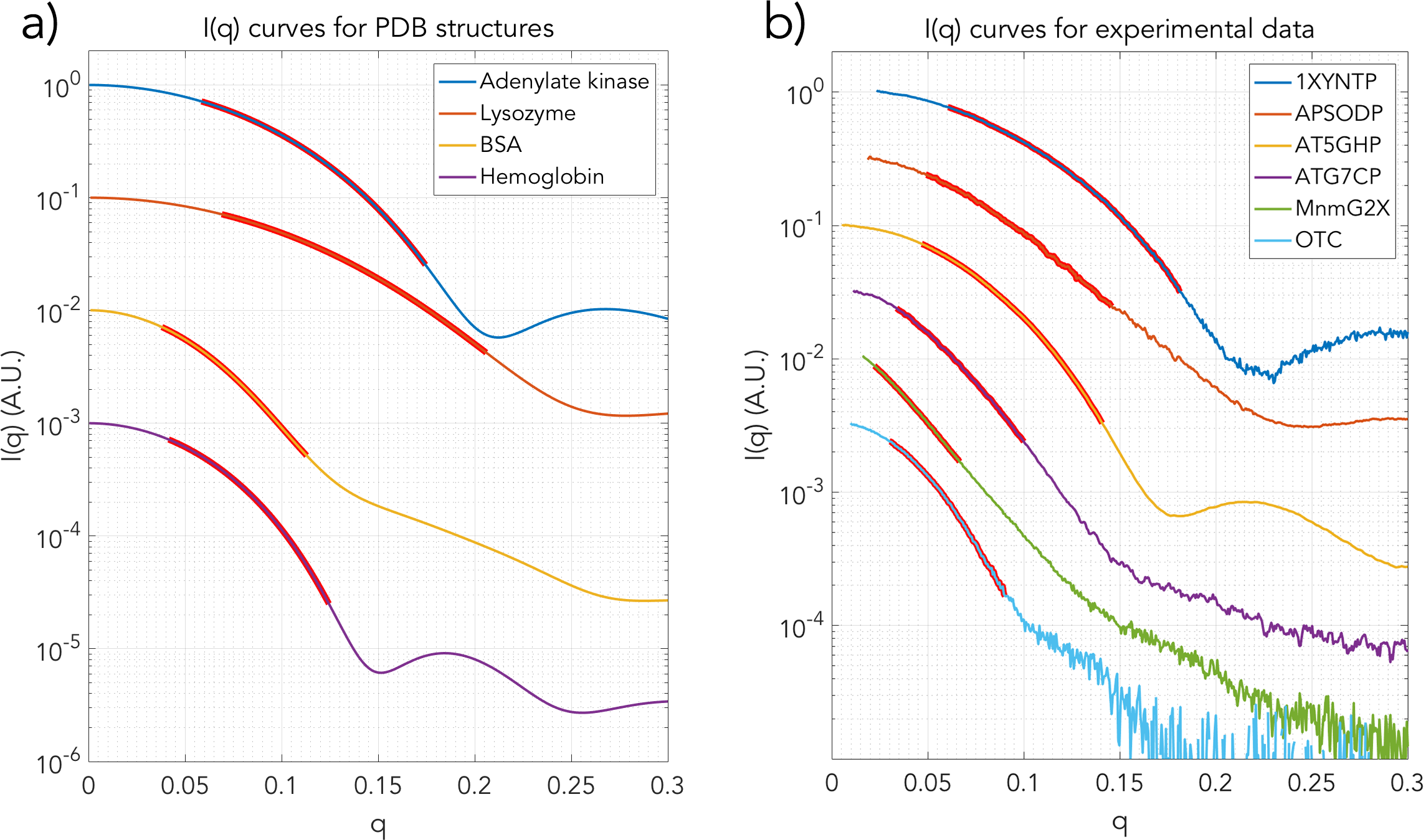}
\caption{(a) SAXS curves for the PDB test cases computed by FoXS \cite{Schneidman-Duhovny2010}. Red background behind each curve shows the $q$-range for the polynomial fit. (b) SAXS curves for the experimental data cases. Red background behind each curve shows the $q$-range for the polynomial fit. }
\label{Fig:Curves}
\end{figure}

\begin{figure}
\includegraphics[width=15cm]{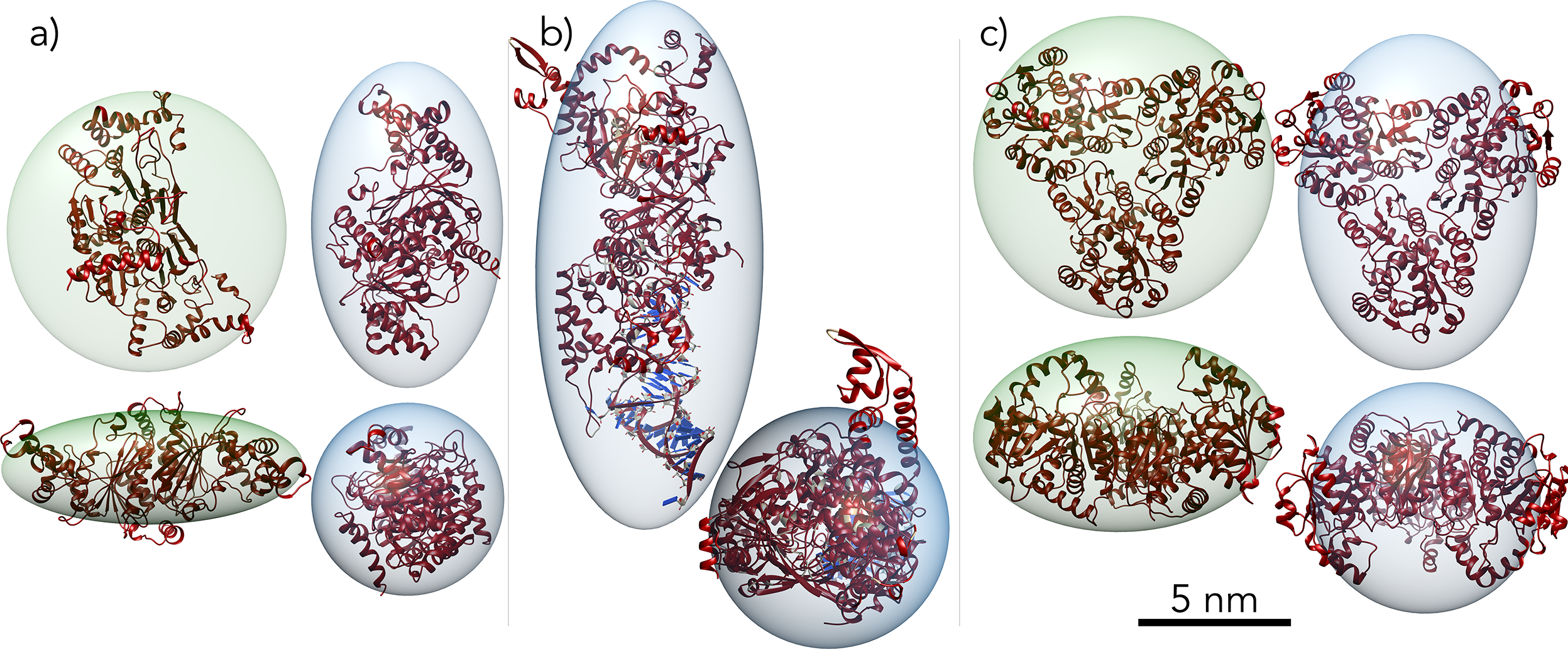}
\caption{Estimated ellipsoidal envelopes for the real data curves of (a) ubiquitin-like modifier-activating enzyme ATG7 (BID ATG7CP, \cite{Taherbhoy2011}), (b) MnmG-tRNA complex (BID MnmG2X \cite{Fislage2014}) and (c) ornithine transcarbamylase (OTC, PDB code 1AKM \cite{Jin1997} and SAXS data collected by our laboratory \cite{Ngu2018}). Color of the ellipsoid indicates the oblate (green) and prolate (blue) candidate. Ellipsoids were generated using the axial lengths provided by our method and manually aligned to the molecular structures rendered with Chimera \cite{Pettersen2004}.}
\label{Fig:ResultsRealData2}
\end{figure}

Next we applied the proposed method to scattering curves of biomolecules. We start with scattering patterns from computational models from the PDB database \cite{Berman2000}. The structures used are those of adenylate kinase (PDB code 1AKE \cite{Muller1992}), lysozyme (PDB code 2LYZ \cite{Diamond1974}), bovine serum albumin (BSA, PDB code 3V03  º\cite{Majorek2012}) and hemoglobin (PDB code 1A3N, \cite{Tame2000}). The scattering curves for each model, shown in Figure \ref{Fig:Curves}.a, were computed with the software FoXS \cite{Schneidman-Duhovny2010}.\\
\\
Given that the proposed approximation uses higher order terms of $I(q)$ compared to the Guinier approximation, the $q$-range in which to carry out the polynomial fit should be expected to contain higher $q$ values than the $q$-range usually employed for the Guinier plot, i.e. $q\,R_G\leq 1.3$. In order to select an appropriate $q$-range, the angle $q$ must be kept low enough to preserve the validity of the proposed approximation, but not too low so as to lose the high order information from which to extract the anisotropy parameters. Experimentally, the range $1\leq q\,R_G\leq 3$ has been found to be a successful trade-off in that regard. To find the appropriate $q$-range for each curve, the radius of gyration $R_G$ for each model was estimated from its corresponding curves using the \textit{Guinier fit} routine from the software suite BioXTAS RAW \cite{Nielsen2009} and the polynomial fit was carried out in the range $1\leq q\,R_G\leq 3$, based on the estimate of $R_G$. The resulting intervals are shown as the red background in the curves shown in Figure \ref{Fig:Curves}.a .\\ 
\begin{figure}
\includegraphics[width=15cm]{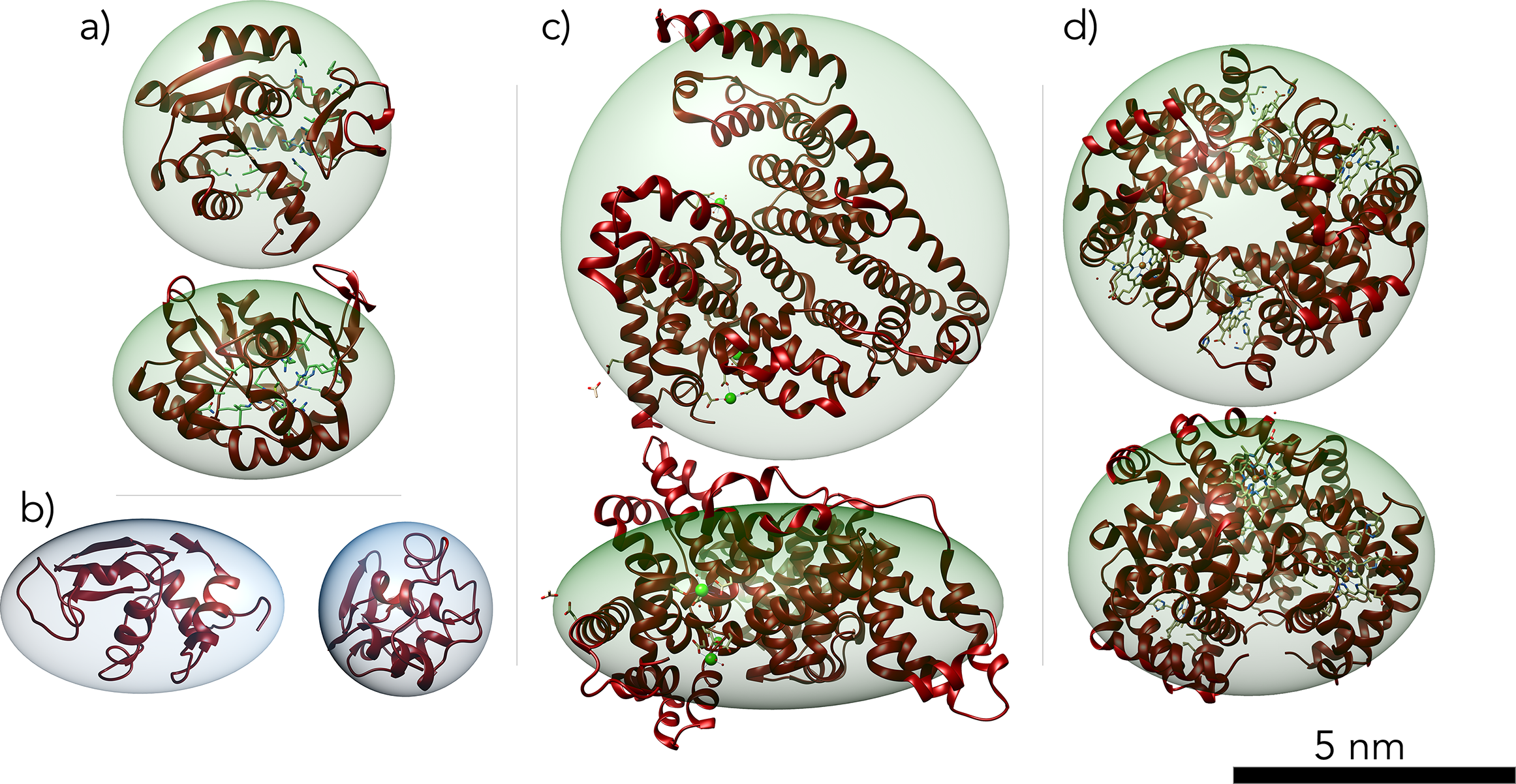}
\caption{Estimated ellipsoidal envelopes for the PDB structures (a) adenylate kinase (PDB code 1AKE \cite{Muller1992}), (b) lysozyme (PDB code 2LYZ \cite{Diamond1974}), (c) bovine serum albumin (BSA, PDB code 3V03 \cite{Majorek2012}) and (d) hemoglobin (PDB code 1A3N \cite{Tame2000}). Color of the ellipsoid indicates the chosen oblate (green) or prolate (blue) candidate. Ellipsoids were generated using the axial lengths provided by our method and manually aligned to the molecular structures rendered with Chimera \cite{Pettersen2004}.}
\label{Fig:ResultsPDBcase}
\end{figure}
\\
Again the polynomial estimation was carried out using a polynomial order of $P = 6$ and the oblate/prolate candidate is chosen by selecting the one that better approximated the $q^6$ fitting term. Figure \ref{Fig:ResultsPDBcase} shows the estimated ellipsoid from each curve, where the color indicates if the chosen ellipsoid is oblate (green) or prolate (blue). The estimated ellipsoids were generated using the axial lengths provided by our method and manually aligned to the molecular structure using the visualization software Chimera \cite{Pettersen2004}. The results illustrate that the proposed approach is able to reliably estimate the oblate/prolateness of the structure and estimate its approximate envelope, even when the structure greatly deviates from an ellipsoidal shape (e.g. BSA in Figure \ref{Fig:ResultsPDBcase}.c).\\
\\
Lastly, we carried out the anisotropy estimation on experimental SAXS curves. We used curves from the BIOISIS database \cite{Hura2009} for which a corresponding PDB model was available, as well as SAXS data collected by our group. The SAXS data used corresponds to xylanase (BID 1XYNTP \cite{Rambo2013}), superoxide dismutase (BID APSODP, \cite{Shin2009}), glycosil hydrolase (BID AT5GHP \cite{Hammel2005}), ubiquitin-like modifier-activating enzyme ATG7 (BID ATG7CP, \cite{Taherbhoy2011}), MnmG-tRNA complex (BID MnmG2X \cite{Fislage2014}) and ornithine transcarbamylase (OTC, PDB code 1AKM \cite{Jin1997} and SAXS data collected by our laboratory \cite{Ngu2018}),\\
\\
The scattering curves for the 6 cases are shown in Figure \ref{Fig:Curves}.b, with the fitting region again shown as red background. The polynomial fit was again carried out in the range $1\leq q\,R_G\leq 3$, computing the $R_G$ from each curve using RAW. Due to the inherent presence of noise in real experiments, the polynomial order was set to $P=4$, to prevent the contribution of the noise to leak into the higher order terms of the fit. Both oblate and prolate candidates are reported, as in this case the higher order terms can't be used to disambiguate them due to the presence of noise and the low order of the fit.\\
\begin{figure}
\includegraphics[width=15cm]{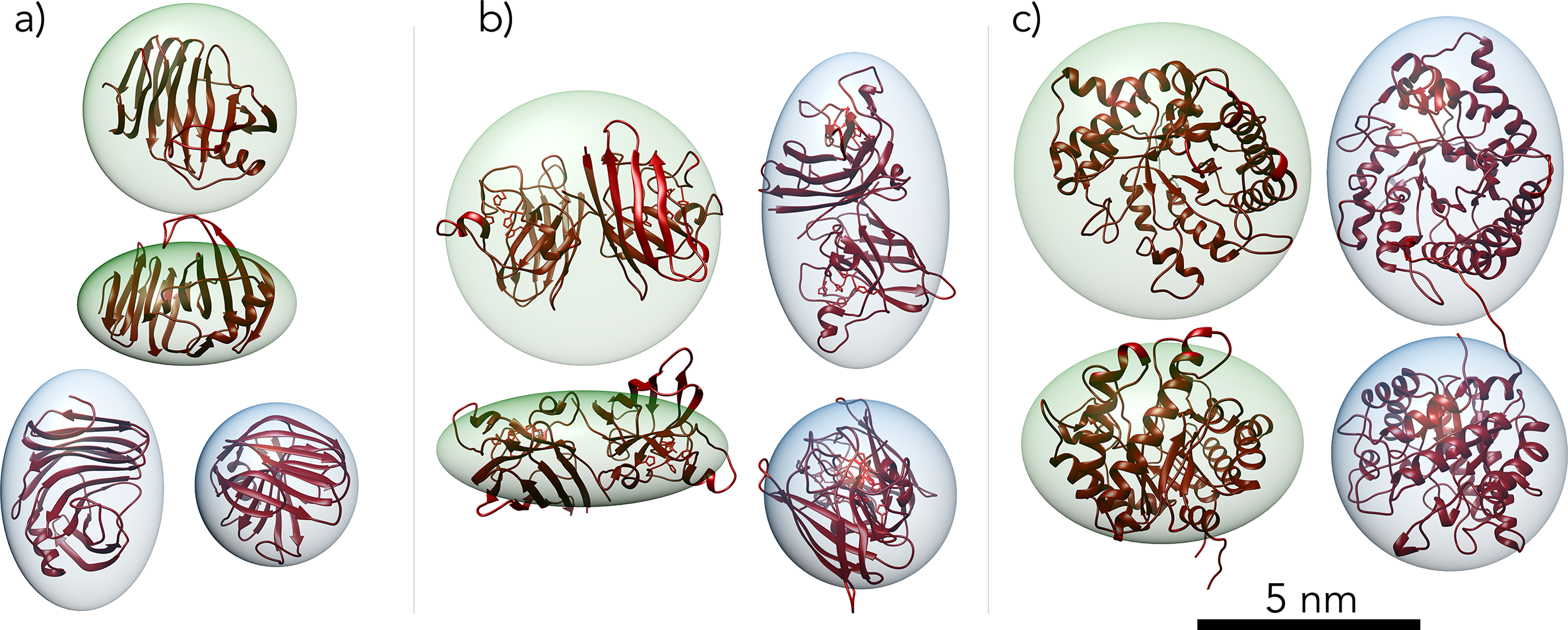}
\caption{Estimated ellipsoidal envelopes for the real data curves of (a) xylanase (BID 1XYNTP \cite{Rambo2013}), (b) superoxide dismutase (BID APSODP, \cite{Shin2009}) and (c) glycosil hydrolase (BID AT5GHP \cite{Hammel2005}). Color of the ellipsoid indicates the oblate (green) and prolate (blue) candidate. Ellipsoids were generated using the axial lengths provided by our method and manually aligned to the molecular structures rendered with Chimera \cite{Pettersen2004}.}
\label{Fig:ResultsRealData1}
\end{figure}
\\
The oblate (green) and prolate (blue) envelope estimates are shown in Figures \ref{Fig:ResultsRealData1} and \ref{Fig:ResultsRealData2}. In the case of MnmG2X (Figure \ref{Fig:ResultsRealData2}.b), the high value value of the anisotropy factor $A_F^2$ yielded a complex $\epsilon$ for the oblate candidate, so only the prolate estimate is shown. The estimated ellipsoid parameters for all test cases are listed in Table \ref{Table:ResultsTable}. Judging from the visual agreement between the molecular structures and the superimposed ellipsoids in Figures \ref{Fig:ResultsRealData1} and \ref{Fig:ResultsRealData2}, in all test cases the presented method is able to provide a satisfactory envelope for at least one of the prolate/oblate candidates. \\

\begin{table}
\centering
 \begin{tabular}{|c|| c | c|} 
 \hline
 Case & Oblate Estimate & Prolate Estimate  \\ [0.5ex] 
 \hline\hline
 1AKE&\textbf{23.96      23.96      17.16}&19.69      19.69      25.82\\
\hline
2LYZ &21.88      21.88      9.937&\textbf{15.01      15.01      24.61}\\
\hline
3V03&\textbf{39.95      39.95      18.33}&27.5       27.5       44.9\\
\hline
1A3N&\textbf{33.17      33.17       25.1}&28.05      28.05      35.45\\
\hline
1XYNTP&24.54      24.54      13.77&18.09      18.09       27.2\\
\hline
APSODP&31.37      31.37      12.53&20.8       20.8      35.49\\
\hline
AT5GHP&30.43      30.43      21.44&24.8       24.8      32.88\\
\hline
ATG7CP&47.45      47.45      17.58&30.93      30.93      53.84\\
\hline
MnmG2X&N/A N/A N/A&35.68      35.68      84.76\\
\hline
OTC &47.26      47.26      30.59&36.99      36.99      51.63\\
\hline
\end{tabular}
\label{Table:ResultsTable}
\caption{Estimated ellipsoid parameters for all test cases. For PDB test cases, bold font indicates the prolate/oblate candidate chosen by the proposed method.}
\end{table}

Given our results, a key topic for future work would be to develop further tools to decide the most suitable candidate between the oblate and prolate alternatives. In that regard, there are several approaches to be investigated. Since Porod's law and the 'power law regime' \cite{Putnam2007} has been claimed to be effective in discerning between oblate and prolate scattering by analyzing the decay of $I(q)$ at high $q$-range, this is one option that could be pursued.\\
\\
Visual inspection of the low $q$ decay of the SAXS curves can also be useful in guiding that decision. We note that for the most elongated real data cases, i.e. APSODP in Figure \ref{Fig:ResultsRealData1}.b, ATG7CP in Figure \ref{Fig:ResultsRealData2}.a and MnmG2X in Figure \ref{Fig:ResultsRealData2}.b, their SAXS curves in Figure \ref{Fig:Curves}.b follow an almost-linear decay in the fitting region (highlighted by the red background), while the other test cases exhibit a more concave decay. This phenomena is due to the contribution of $A_F^2$ in Equation \ref{FinalApprox}: if $A_F^2$ is low, the curve is dominated by the negative quadratic term, creating a concave decay in the SAXS curve. However, as $A_F^2$ increases, the convex trend in the quartic term compensates for the concave quadratic, resulting in the approximately linear decay.\\
\\
An additional criteria might be developed by considerations of the mechanics of protein folding/stability to assess which degrees of oblate/prolateness are most likely, or even possible. For the most elongated test cases mentioned above, the oblate $\epsilon^2$ is below 0.2, which results in a very flat oblate envelope that appears to be a very rare protein shape, perhaps largely precluded by considerations of stability. In those cases, the prolate candidate would appear the most likely.\\
\\
One concern with our approach to estimate  $R_G$ and $A_F$ is that the simple polynomial fit $I(q) \approx \sum_i^P a_i q^{2i}$ in Section \ref{Section:Methods} ignores certain constraints imposed by the model in Equation \ref{FinalApprox}. For example, we ignore the fact that, as $R_G^2$ is  non-negative,  the coefficient $a_1$ associated with $q^2$ in the polynomial  must be non-positive, since $a_1 \approx - \frac{R_G^2}{3}$. (This is also true for the Guinier plot approach, but since  SAXS curves generally have a negative slope near the origin, $a_1$ is naturally negative without the need to explicitely force it to be.)\\
\\
Our model poses an additional constraint: given that $a_2 \approx \frac{1}{21} \left( R_G^4 + \frac{1}{5}  A_F^2 \right)$ and that $A_F^2$ is non-negative, we have that $\frac{1}{21} R_G^4 \leq \frac{1}{21} \left( R_G^4 + \frac{1}{5}  A_F^2 \right)$, leading to the constraint $\frac{9}{21} a_1^2 \leq a_2$. Not enforcing this constraint in the polynomial fit could lead to negative estimates of $A_F^2$, in conflict with its physical meaning. This concern could be addressed by instead performing a constrained polynomial fit:\\
\\
\begin{equation}
\begin{split}
    \operatorname*{minimize}_a\quad &||I - V a||_2^2 \\
    \text{subject to} \quad &\frac{9}{21} a_1^2 \leq a_2
    \end{split}
    \label{optiproblem}
\end{equation}
\\
where $I \in R^N$ is the SAXS intensity pattern, $a \in R^{P+1}$ are the coefficients of the polynomial fit and given $q = [q_1,\dots,q_N]$, the associated scattering angles for $I$, $V$ is the Vandermonde matrix of even powers of $q$:
\begin{equation}
    V = \begin{bmatrix}
      1 & q_1^2 & q_1^4 & \dots & q_1^{2P}\\ 
       1 & q_2^2 & q_2^4 & \dots & q_2^{2P}\\
       \vdots & \vdots & \vdots & \dots & \vdots\\
       1 & q_N^2 & q_N^4 & \dots & q_N^{2P}\\
     \end{bmatrix}
     \label{matrixQ}
\end{equation}
The unconstrained polynomial fit can be seen as the solution of the optimization problem in Equation \ref{optiproblem} without the inequality constraint and can easily be solved in closed form. The constrained version, however, is an instance of a more general class of problems known as quadratically constrained quadratic problems (QCQP) \cite{Boyd2004} and can generally only be solved by iterative methods. However, for the task of estimating a scatterer's anisotropy, we show next that the computationally simpler unconstrained fit is enough to recover $A_F^2$.\\
\\
To do so we rely in the convexity of \ref{optiproblem}. When solving the unconstrained problem, the optimal polynomial parameters $a$ can either satisfy or violate the inequality constraint in \ref{optiproblem}. If the constraint is satisfied, then we are guaranteed that $a$ are also the optimal parameters for the constrained problem. If the constraint is violated, by the convexity of \ref{optiproblem}, we are guaranteed that the optimal constrained solution will lie on the boundary of the feasible region, i.e. will satisfy the inequality constrained with equality: $\frac{9}{21} a_1^2 = a_2$. When this happens, we have that $A_F^2 = 0$ and $\epsilon = 1$, characteristic of a spherical scatterer. We can then rely on solving just the unconstrained problem and applying following guideline: if the resulting $A_F^2$ is positive, we keep it; if it is not, then $A_F^2 = 0$. 


\subsection{Effect of particle anisotropy on the estimate of $R_G$}
\begin{figure}
\includegraphics[width=8cm]{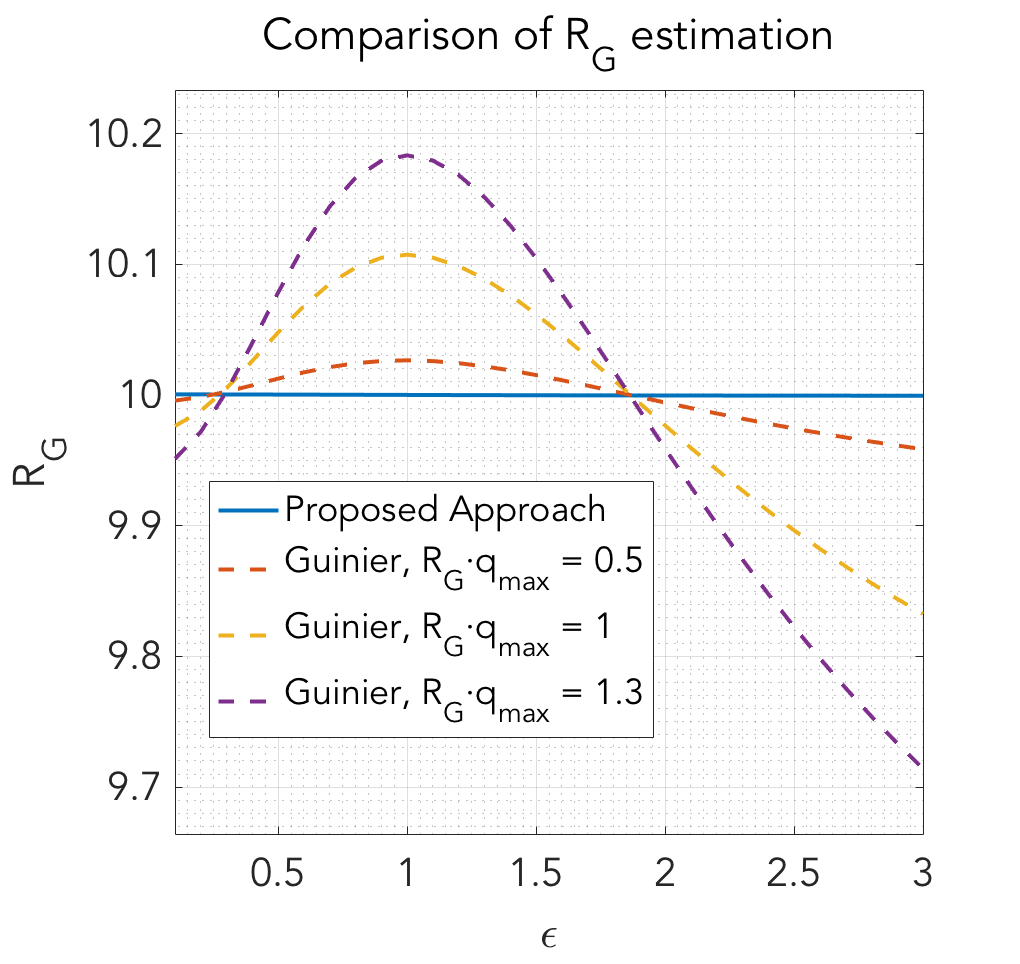}
\caption{Comparison of $R_G$ estimation between proposed approximation and Guinier approximation for set of ellipsoids with $R_G = 10$ and $\epsilon = [0.1,\dots,3]$. The proposed approach correctly estimates the $R_G$ in all cases, while the Guinier approximation presents a non-linear deviation from the true $R_G$ consistent with Equation \ref{Eq:ModelMismatch}.}
\label{Fig:Rg_comp}
\end{figure}

Finally, we analyze the the accuracy of our method's $R_G$ estimation and compare it to that of Guinier's. While the goal of the proposed method is not to more accurately estimate $R_G$ but to infer the anisotropy characteristics of the scatterer, we find it of interest to analyze how much both approaches deviate from one another in their $R_G$ estimation. To get a quantitative measure of their difference, we calculate the discrepancy between both approximations. The scattering intensity in the Guinier case, denoted by $I_G$, can be deduced by taking the first 3 terms of the Taylor series of Equation \ref{GuinierAppox}:
\begin{equation}
   I_G(q) =  1 - \frac{R_G^2 q^2}{3} + \frac{R_G^4 q^4}{18} + \mathcal{O}(q^6)
\end{equation}
Aside from the lack of the anisotropy factor from Equation \ref{FinalApprox}, $I_G$ gives a slightly greater importance to the $q^4$ term, where $R_G^4$ is scaled by $1/18$, instead of the $1/21$ factor in our approximation. The model mismatch is given by:
\begin{equation}
\begin{split}
  \xi(\epsilon) &= I_R - I_G \\
  &= \left(\left(\frac{1}{21}-\frac{1}{18}\right)  R_G^4 + \frac{1}{105}  A_F^2 \right)\, q^4 + \mathcal{O}(q^6) \\
&= 
\frac{19\,R_G^4}{630}\frac{\left(\epsilon^2 - 3.5191\right)\left(\epsilon^2 - 0.0598\right)}{\left(2+\epsilon^2\right)^2}\,q^4  + \mathcal{O}(q^6)  
\end{split}
\label{Eq:ModelMismatch}
\end{equation}
The discrepancy function $\xi(\epsilon)$ is zero at $\epsilon = \sqrt{3.5191} = 1.876$ and $\epsilon = \sqrt{0.0598} = 0.245$, negative between these roots, and positive everywhere else. The expected effect on the $R_G$ estimation from this discrepancy is that the Guinier approximation would tend to underestimate $R_G$ wherever $\xi(\epsilon)$ is positive and to overestimate $R_G$ when $\xi(\epsilon)$ is negative. To test this hypothesis, we estimated the $R_G$ of the 30 ideal ellipsoids presented in Section 3 by carrying out a linear fit to their Guinier plot in the range $[q_{min}, q_{max}]$ for $q_{min}\,R_G =0$ and $q_{max}\,R_G= 0.5, 1$ and $1.3$, as well as estimating $R_G$ using the proposed approach. Figure \ref{Fig:Rg_comp} shows the resulting $R_G$ estimates.\\
\\
Our approach correctly estimated $R_G$ in all cases. The Guinier plot provided a perfect $R_G$ estimate for $\epsilon \approx 0.25$ and $\epsilon \approx 1.85$, coinciding with the approximate location of the zeros of Equation \ref{Eq:ModelMismatch}, and exhibited the under/overestimation pattern predicted by $\xi(\epsilon)$, with the maximum overestimation given by the spherical case $\epsilon = 1$. The Guinier results also showed a $q_{max}$ dependency, where the disagreement between the Guinier estimation and the true $R_G$ increases with $q_{max}$, due to $q^4$ term of the discrepancy $\xi(\epsilon)$.\\
\\
However, the magnitude of the Guinier deviation from the correct $R_G$ is at most 3 percent points in Figure \ref{Fig:Rg_comp}, while the discrepancy between the $R_G$ estimated by our method and Guinier's is at most 5 percent points for the computed and experimental SAXS curves presented in this section, demonstrating the strength of the Guinier approximation in reliably estimating $R_G$ for a variety of scattering geometries. The proposed approximation should be used, we think, not in substitution of the Guinier plot, but as a complement to it, providing an additional estimate on the scatterer's anisotropy, as well as a refinement on the $R_G$, assuming sufficiently low noise levels.\\

%% file: Tex/Conclusions.tex
In this work we have developed an approximation for SAXS curves of ellipsoids of revolution that allows for an estimation of the anisotropy of the scatterer. This approximation is shown to be reasonably accurate even for non-ellipsoidal scatterers, accurately estimating the anisotropic envelope of molecules from their SAXS curves, both computed from their PDB models and from experimental data. The proposed approach can be useful in SAXS data analysis as an extension of the Guinier approximation and for a more data-driven initialization of \textit{ab initio} structure reconstruction algorithms. \\
\\